\documentstyle[eqsecnum,aps,epsf]{revtex}
\begin{document}

\draft

\newcommand {\beq}{\begin{eqnarray}}
\newcommand {\eeq}{\end{eqnarray}}
\newcommand {\be}{\begin{equation}}
\newcommand {\ee}{\end{equation}}
\newcommand{\Gmu}{\gamma^{\mu}}
\newcommand{\Gnu}{\gamma^{\nu}}
\newcommand{\gmu}{\gamma_{\mu}}
\newcommand{\gnu}{\gamma_{\nu}}
\newcommand{\bg}{\mbox{\boldmath $\gamma$}}
\newcommand{\gfour}{\gamma_4}
\newcommand{\del}{\partial}
\newcommand{\k}{\mbox{\boldmath $k$}}
\newcommand{\q}{\mbox{\boldmath $q$}}
\newcommand{\p}{\mbox{\boldmath $p$}}
\newcommand{\wn}{\omega_n}
\newcommand{\wm}{\omega_m}
\newcommand{\La}{{\Lambda_{\scriptsize{\mbox{QCD}}}}}
\newcommand{\CC}{\langle \bar{q} q \rangle}

\topmargin=0cm

\title{Universality, the QCD critical/tricritical point and the quark number susceptibility}

\author{
$^{(1,2)}$Yoshitaka Hatta and $^{(3)}$Takashi Ikeda
}

\address{
$^{(1)}$Department of Physics, Kyoto University, 
         Kyoto 606-8502, Japan\\
$^{(2)}$The Institute of Physical and Chemical Research (RIKEN), Wako, Saitama 351-0198, Japan\\
$^{(3)}$RIKEN BNL Research Center, Brookhaven
         National Laboratory, Upton, New York 11973-5000, USA
}

\date{\today}

\maketitle

\begin{abstract}
The quark number susceptibility near the QCD critical end-point (CEP), the tricritical point (TCP) and the O(4) critical line at finite temperature and quark chemical potential is investigated.
Based on the universality argument and numerical model calculations we propose a possibility that the hidden tricritical point strongly affects the critical phenomena around the critical end-point.
We made
 a semi-quantitative study of the 
 quark number susceptibility near CEP/TCP for
 several quark masses          
 on the basis of the Cornwall-Jackiw-Tomboulis (CJT) potential for QCD in the improved-ladder
 approximation. The results show that the susceptibility is enhanced in a wide region around CEP inside which the critical exponent gradually changes from that of CEP to that of TCP, indicating a crossover of different universality classes. 

\end{abstract}


\section{INTRODUCTION}

 The vacuum of the quantum chromodynamics (QCD) is believed to
undergo a phase transition to the quark-gluon plasma (QGP) 
at high temperature $T$ and/or at high quark chemical potential $\mu$.
 Such a new state of matter is expected 
to be produced in on-going heavy-ion collision experiments
 at Relativistic Heavy-Ion Collider (RHIC) 
 and in the future Large Hadron Collider (LHC)  \cite{QMproceedings}.

The phase transition of the hadronic matter to QGP  at finite $T$ with $\mu=0$
 has been studied extensively on the lattice. In particular,
 the chiral phase transition 
 is likely to be of  second order 
 for QCD  with two massless quarks. Also, the static critical behavior
 is expected to  fall into 
 the universality class of the O(4) spin model in three dimensions
\cite{rob}.
In nature, the light quarks have small but finite masses
and the second order transition becomes a smooth crossover. 

 Study of the QCD phase transition with finite  $\mu $
 has been retarded because reliable lattice simulations
 have not been available so far due to the severe fermion sign problem.
 Nevertheless, there is a growing evidence that  the
  phase diagram of QCD with  massless 2-flavors has a tricritical point (TCP, Fig. \ref{phase}, point P) at which 
 a line of critical points (the O(4) line)
  at lower $\mu$'s turns into a first order phase transition line
 at  higher $\mu$'s. 
 The existence of TCP was in fact suggested in various 
 calculations based on 
 effective theories of QCD 
\cite{Asakawa:bq,Halasz:1998qr,Berges:1998rc,Harada:1998zq,Klevansky:qe,Barducci:1993bh,Alford:1997zt,Kiriyama:2001ah}.
 If the $u$, $d$-quark masses are increased from zero,  a line of critical points (the wing critical line) emerges from TCP and the point which corresponds to the physical quark mass $m_{phys}$ is called the QCD critical end-point (CEP, Fig. \ref{phase}, point E) because this is the point 
 where  the first order phase transition line  terminates. Indeed, some evidence of the existence of CEP was shown recently  
  in a lattice QCD simulation
 with 2+1 flavors by Fodor and Katz \cite{Fodor:2001au}. In this paper we assume that CEP exists in the phase diagram of QCD.\footnote{Accordingly, we fix the strange quark mass to its physical value.
Below `the quark mass' means the $u$, $d$ current quark mass which we consider as a variable parameter.}  
 
Second order phase transitions are characterized by the long-wavelength fluctuations of the order parameter. In the case of CEP, it is the sigma ($\sigma $) field. Then, it is expected that the fluctuations of the sigma field will be reflected in the event-by-event fluctuation of pion ($\pi $) observables due to the the $\pi -\sigma$ coupling. Based on this observation, possible observable signals associated with CEP have been studied in detail in relation to the relativistic heavy-ion collision experiments \cite{Stephanov:1998dy,Berdnikov:1999ph,fukushima}. 

The purpose of this paper is to point out that the anomaly near CEP is not pointlike but has much richer structure. Our starting point is a simple question: ``How large is the critical region?'' The critical region is defined as the region where the mean field theory (or the Landau theory) of phase transitions breaks down and the true non-trivial critical exponents can be seen. Usually, one expects that the critical region is surrounded by the mean field region and the critical exponents change from the non-trivial values to the mean field values as one comes away from the critical point. One might argue that this question is only of academic interest because the non-trivial exponents and the mean field exponents are numerically not so different and probably experiments cannot distinguish them. [This observation is the basis of \cite{Stephanov:1998dy}.] However, as we will see, pursuing this question leads to an important notion which may shed light on certain results of both heavy-ion collision experiments and future lattice simulations at finite chemical potentials.      

There is a well-known criterion which estimates the size of the critical region, the Ginzburg criterion \cite{gin}. It tells that if the singular part of the thermodynamic potential $\Omega$ (the Landau-Ginzburg potential) for a certain second order phase transition is given by 
\begin{eqnarray}
\Omega|_{singular}=c(\nabla \phi)^2+at\phi^2 +b\phi^4,\label{gg}
\end{eqnarray}
where $\phi$ is the order parameter and $t\equiv \frac{T-T_c}{T_c}$ is the reduced temperature ($T_c$ is the critical temperature in the mean field approximation), the critical region is estimated to be 
\begin{eqnarray}
|t|\sim \frac{T_c^2b^2}{ac^3}.\label{tg}
\end{eqnarray}
At first sight, this criterion seems useless because we do not know the coefficients appearing in (\ref{gg}) for CEP.\footnote{The size of the critical region depends on the microscopic dynamics and universality tells nothing about it. A clear example is the $\lambda$-transition of liquid helium and the superconducting transition of metals. Although they belong to the same universality class (the O(2) spin model), their critical regions are very different; $|t|\sim 0.3$ for the $\lambda$-point and $|t|\sim 10^{-15}$ for conventional (type-I) superconductors. Just for reference, we note that for typical liquid-gas phase transitions which belong to the same universality class as the phase transition at CEP, $|t|\sim 10^{-2}$ \cite{ley}. [Corrections to the scaling \cite{wegner} are not negligible until one reaches $|t|\sim 10^{-4}$.]}  However, in the next section we will derive a bound to the size of the critical region. In fact, there is a reason to expect that the critical region of CEP is {\it small}. This is because the QCD critical end-point is a descendant of the tricritical point of the massless theory.

This observation led us to study the critical phenomena of both CEP and TCP simultaneously and their possible correlations. We make both qualitative and quantitative analyses of the 
 physics near TCP and CEP with particular emphasis
 on the (singular) behavior of the quark number susceptibility $\chi_q$ defined by 
\begin{eqnarray}
\chi_q=-\frac{1}{V}\frac{\partial ^2\Omega}{\partial \mu ^2},
\end{eqnarray}
where $\Omega (T, \mu)$ is the thermodynamic potential and $V$ is the volume of the system.
$\chi_q$ is a response of the quark number
 density to the variation of the quark chemical potential and 
 is one of the key quantities characterizing the phase change from the 
 hadronic matter to QGP \cite{McLerran:1987pz,Kunihiro:1991qu,go,asakawa,blaizot}.
 The lattice data tell us that, at $\mu=0$, $\chi_q$ increases rapidly but smoothly near the critical temperature \cite{Gottlieb:ac,Gavai:1989ce,Gavai:2001fr}.
On the other hand, the universality argument predicts that it diverges at both TCP and CEP with certain critical exponents.
  Therefore, it would be 
 important to study its critical behavior with and without
 the quark masses to see whether it can provide a 
 new way of detecting the TCP/CEP on the lattice as well as 
 in the heavy-ion collision experiments.

In addition to $\chi_q$, we occasionally mention the singular behavior of the specific heat $C$ and the chiral susceptibility $\chi_{ch}$ defined as 
\begin{eqnarray}
C=-\frac{T}{V}\frac{\partial ^2\Omega}{\partial T^2},\\
\chi_{ch}=\frac{1}{V}\frac{\partial ^2\Omega}{\partial m^2}
\end{eqnarray}
From the viewpoint of critical phenomena, $\chi_q$ and $C$ are essentially the same near TCP/CEP while $\chi_{ch}$ is different from $\chi_q$ near TCP and only slightly different near CEP in the sense that will be clarified below.

In Section II, we will make a general analysis of the 
 interplay between TCP and CEP in the small quark mass limit based on the universality argument.
 After determining the
  relative location of TCP and CEP in the phase diagram
 as a function of the quark mass, we construct the Landau-Ginzburg potential for CEP to determine the singular behavior of susceptibilities both in and beyond the mean field approximation. It turns out that the smallness of the quark mass gives a bound to the growth of the critical region, turning our attention to the tricritical point. Then we discuss a possible crossover from the tricritical universality class to the Ising universality class.

The universality argument is so general that it gives no quantitative results. In order to reinforce the ideas given in Section II, in Section III we show the results of the numerical calculation on a model, the Cornwall-Jackiw-Tomboulis (CJT) potential for QCD \cite{CJT} in the improved ladder approximation \cite{Kiriyama:2001ah,Kiriyama:2001nq}. We will find how well the numerical results match with the qualitative predictions of the universality argument, demonstrating the power of universality. In particular, we observed some indication of the effect of TCP on the QCD phase diagram even with a reasonable value of the quark mass. 

Section IV is devoted to conclusions.

Brief description of the model is given in Appendix A.
In Appendix B, we discuss, for completeness, how $\chi_q$ behaves along the O(4) line based on the universality argument. We will see that the monotonous increase so far observed on the lattice is a property only at $\mu=0$.

\section{Universality arguments}
Universality is such a strong notion of modern physics \cite{ma} that its applicability ranges from phase transitions in ordinary liquids to thermal phase transitions of relativistic quantum field theories. In this section we study the critical phenomena near CEP/TCP based on the universality argument. We will see that a lot of general information can be extracted by the universality argument alone without mentioning any complexities of the strong interaction. 

\subsection{The QCD critical end-point}
\begin{figure}[htbp]
    \centerline{
      \epsfxsize=0.49\textwidth
      \epsfbox{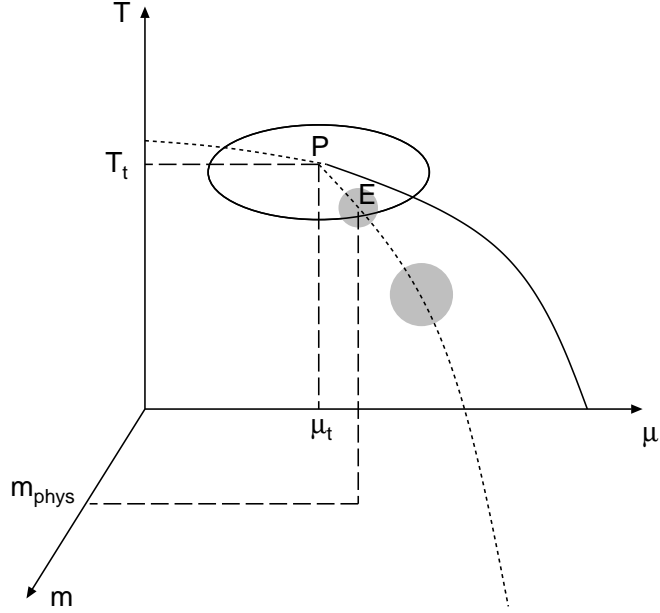}
    }
  \caption{The phase diagram of QCD in the $(T,\mu,m)$ space. Point P is the tricritical point (TCP) of the massless theory and point E is the critical end-point (CEP) of the real world. The dotted lines represent the second order phase transition and the solid line represents the first order phase transition.
           }
  \label{phase}
\end{figure}

It was suggested theoretically \cite{Halasz:1998qr,Berges:1998rc,Barducci:1993bh,Kiriyama:2001ah} and found on the lattice \cite{Fodor:2001au} that QCD has the critical end-point (CEP) at finite temperature $T_c$ and baryon chemical potential $\mu_c$ (Fig. \ref{phase}, point E). At the critical end-point, only the $\sigma$-field becomes massless and the universality class of this phase transition is considered to be the same as that of the liquid-gas phase transition, or equivalently, that of the 3-dimensional Ising model.\footnote{This is not obvious {\it a priori} and requires explanation. As we shall see below, the phase transition at the end-point is characterized by the one-component order parameter $\hat{\sigma}$. The effective Landau-Ginzburg potential contains odd powers of $\hat{\sigma}$ which break the $\hat{\sigma}\to -\hat{\sigma}$ symmetry of the Ising model. This is the same situation as the liquid-gas phase transition. Theoretically, the usual renormalization group argument should be reconsidered in the presence of the asymmetry \cite{hubbard}. Although there are some subtleties about this problem, {\it experimentally} it is clear that the liquid-gas phase transition and the 3D Ising model belong to the same universality class.}
 
In order to exploit the power of universality to investigate the singular behavior of various quantities, we consider the mapping of the $(t_I,h_I)$ axes of the Ising model ($t_I$ is the reduced temperature and $h_I$ is the reduced magnetic field) onto the $(T,\mu,m)$ space ($m$ is the light quark mass divided by the typical scale of the problem such as $T_c$). This can be achieved by considering the tricritical point (TCP) at $(T,\mu,m)=(T_t,\mu_t,0)$ (Fig. \ref{phase}, point P). Below we explicitly construct the Landau-Ginzburg potential for CEP starting from the general theory of tricritical points \cite{lawrie} and discuss associated universal behaviors.

Near TCP, long-wavelength physics of the system can be described by the thermodynamic potential expanded up to the sixth-order in the order parameter field (the sigma field) $\sigma$
\begin{eqnarray}
\Omega_{MF}=\Omega_0-m\sigma+\frac{a}{2}\sigma^2+\frac{b}{4}\sigma^4+\frac{c}{6}\sigma^6,\label{gl}
\end{eqnarray}
where $\Omega_0$ is the contribution from short wavelength degrees of freedom irrelevant to the study of critical phenomena.  

At TCP, $a=b=m=0$. Assuming that $a$ and $b$ are analytic in $T$ and $\mu$ and that $c>0$ is approximately constant near the tricritical point, we expand them as follows \cite{valid}
\begin{eqnarray}
a(T, \mu)&=&C_a(T-T_t)+D_a(\mu-\mu_t) \nonumber \\
b(T,\mu)&=&C_b(T-T_t)+D_b(\mu-\mu_t),\label{ab}
\end{eqnarray}
where we have neglected terms higher order in the deviation from the tricritical point. $C_a>0$ and $D_a>0$ are related such that the line $a(T,\mu)=0$ is tangential to the first order phase transition curve at TCP. $b$ is positive for $T-T_t>0$ ($\mu<\mu_t$) on the $a=0$ line, which leads to the condition
\begin{eqnarray}
C_bD_a-C_aD_b>0.\label{con}
\end{eqnarray} These conditions come from the geometry of the phase diagram, namely, the fact that there is a line of (bi-)critical points at $T>T_t$, $\mu<\mu_t$. We do not know the actual values of these coefficients. But we need not know them for the present purpose.

   If we increase $m$ from zero, at some point $(T_c(m),\mu_c(m))$ in the $(T, \mu)$ plane two minima and a maximum of the potential coalesce. This is the critical end-point. There the sigma field acquires a non-zero expectation value $\sigma_0$ which is determined by the following equations [In this section we exclusively consider the small $m$ limit and leave only the leading terms in $m$.]    
\begin{eqnarray}
\Omega' (T_c(m),\mu_c(m),\sigma_0)&=&-m+a_m\sigma_0+b_m\sigma_0^3+c\sigma_0^5=0\nonumber \\
\Omega'' (T_c(m),\mu_c(m),\sigma_0)&=&a_m+3b_m\sigma_0^2+5c\sigma_0^4=0\nonumber \\
\Omega''' (T_c(m),\mu_c(m),\sigma_0)&=&6b_m\sigma_0+20c\sigma_0^3=0,
\end{eqnarray}
where $a_m\equiv a(T_c(m),\mu_c(m))$ and $b_m\equiv (T_c(m),\mu_c(m))$.
The solution is 
\begin{eqnarray}
a_m&=&\frac{9b_m^2}{20c}, \nonumber \\
-b_m&=&\frac{5}{54^{\frac{1}{5}}}c^{\frac{3}{5}}m^{\frac{2}{5}},\nonumber \\
\sigma_0&=&\sqrt{\frac{-3b_m}{10c}}.\label{sol}
\end{eqnarray}
Using (\ref{ab}) and (\ref{sol}) we can locate the critical end-point
for small $m$.
\begin{eqnarray}
T_c(m)-T_t&=&-\frac{45D_ac^{\frac{1}{5}}}{4(54)^{\frac{2}{5}}(C_bD_a-C_aD_b)}m^{\frac{2}{5}}+O(m^{\frac{4}{5}}),\nonumber \\
\mu_c(m)-\mu_t&=&\frac{5C_ac^{\frac{3}{5}}}{(54)^{\frac{1}{5}}(C_bD_a-C_aD_b)}m^{\frac{2}{5}}+O(m^{\frac{4}{5}}).\label{mm}
\end{eqnarray}
Thus, as we increase the quark mass $m$, the critical temperature decreases and the critical chemical potential increases at least for small $m$ [Fig. \ref{phase}].
Expanding $\Omega(T,\mu,\sigma)$ around $(T_c(m),\mu_c(m),\sigma_0)$ we obtain the Landau-Ginzburg potential with the new order parameter $\hat{\sigma}\equiv \sigma-\sigma_0$
\begin{eqnarray}
\Omega_{MF}(T,\mu,\hat{\sigma})&=&\Omega_{MF}(T_c(m),\mu_c(m),\sigma_0)\nonumber \\
&+&A_1\hat{\sigma}+A_2\hat{\sigma}^2+A_3\hat{\sigma}^3+A_4\hat{\sigma}^4,\label{33}
\end{eqnarray}
where
\begin{eqnarray}
A_1&=&(C_a\sigma_0+C_b\sigma_0^3)(T-T_c(m))\nonumber \\
&&+(D_a\sigma_0+D_b\sigma_0^3)(\mu-\mu_c(m)),\nonumber \\
A_2&=&(C_a+3C_b\sigma_0^2)(T-T_c(m))\nonumber \\
&&+(D_a+3D_b\sigma_0^2)(\mu-\mu_c(m)), \nonumber \\
A_3&=&C_b(T-T_c(m))+D_b(\mu-\mu_c(m)),\nonumber \\
A_4&=&\frac{-b_m}{2}.\label{44}
\end{eqnarray}
$A_i\ (i=1,2,3)$ vanish at the critical point whereas $A_4$ does not, indicating that $(T_c(m),\mu_c(m))$ is an ordinary (bi-)critical point as stated above.

\begin{figure}[htbp]
    \centerline{
      \epsfxsize=0.49\textwidth
      \epsfbox{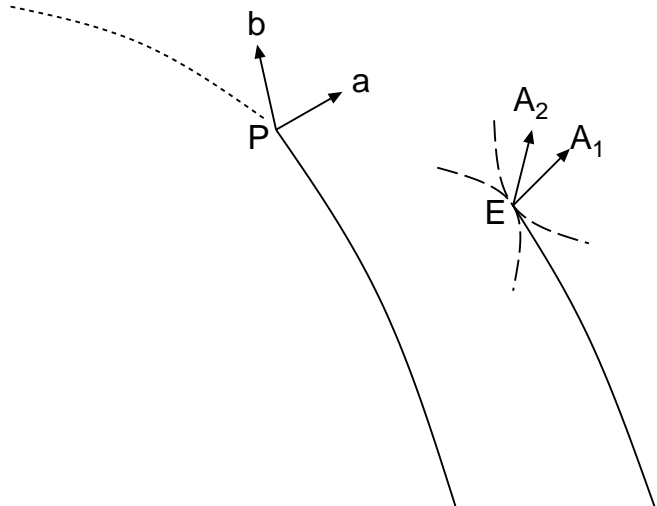}
    }
  \caption{The mapping of the Ising model axes onto the $(T,\mu)$ plane. The solid lines represent the first order phase transition (the coexisting line). The dashed lines separate regions with different exponents.
           }
  \label{map}
\end{figure}

Looking at (\ref{33}) and (\ref{44}), we immediately notice two important things. First, $A_2$ is not proportional to $T-T_c(m)$ as in the Landau theory but a linear combination of $T-T_c(m)$ and $\mu-\mu_c(m)$. This means that $T$ and $\mu$ are equivalent thermodynamic variables in the sense of Griffiths and Wheeler \cite{griffiths} and that $A_2$ is the temperature-like scaling field which corresponds to $t_I$ of the Ising model. Second, $A_1$ rather than the quark mass plays the role of the `external field' which is conjugate to the new order parameter. Thus it can be identified as the magnetic field-like scaling field $h_I$. Indeed, it is easy to show that, on the line $A_1=0$, $A_2$ and $A_3$ are positive for $T>T_c(m)$ (or $\mu<\mu_c(m)$) and negative for $T<T_c(m)$ (or $\mu<\mu_c(m)$) and this line is asymptotically parallel to the first order phase transition line at the critical end-point. See Fig. \ref{map}.

Now we can discuss the critical behavior of susceptibilities; the quark number susceptibility $\chi_q$, the specific heat $C$ and the chiral susceptibility $\chi_{ch}$. In the mean field approximation, the equilibrium value of $\hat{\sigma}$ is determined by the first and fourth order terms of (\ref{33}) in the small mass limit. Then we obtain, for paths asymptotically not parallel to the $A_1=0$ line (the first order phase transition line)
\begin{eqnarray}
\chi_q&\sim &C\sim m^{\frac{2}{15}}|g-g_c|^{-\epsilon},\nonumber \\
\chi_{ch}&\sim &m^{-\frac{16}{15}}|g-g_c|^{-\epsilon} \label{cc},
\end{eqnarray} 
where $\epsilon \equiv \frac{\gamma}{\beta \delta}=\frac{2}{3}$. $|g-g_c|$ denotes the distance from CEP in some units. For the path asymptotically parallel to the $A_1=0$ line, the exponent is $\gamma=1 >\epsilon$. Note that, although the critical exponents are the same, the {\it amplitude} of the chiral susceptibility is enhanced whereas that of the quark number susceptibility is suppressed by factors of $m$.

 Inside the critical region, where the mean field theory breaks down, $\Omega|_{singular}$ does not admit a simple expansion with smooth coefficients.
(\ref{33}) should be regarded as the saddle point approximation to the following functional integral
\begin{eqnarray}
\Omega(T,\mu,\hat{\sigma})=-\frac{T}{V}\ln \int [d\sigma' ]\exp\bigl(-\frac{1}{T}\int d^3 {\bf r}H_{eff}({\bf r})\bigr),\label{true}
\end{eqnarray}
where $H_{eff}$ is the Landau-Ginzburg-Wilson hamiltonian
\begin{eqnarray}
H_{eff}=A'_0(\nabla \sigma')^2+A'_1\sigma '+A'_2\sigma '^2+A'_3\sigma'^3+A'_4\sigma '^4.\label{eff}
\end{eqnarray}
$A'_i\ (i=1\sim 4)$ are in general different from $A_i$ due to fluctuations. However, we expect that the differences between $A'_i$ and $A_i$ are of the higher order in $m$. \footnote{The coefficients are further affected by the change of integration variables.  These degrees of freedom can eliminate $A'_3$, but do not change $A'_{1,2}$ in the leading order. In fact, only the direction of $A'_1$ is important to discuss the behaviors of quantities considered here (i.e., second derivatives of $\Omega$ in directions parallel to the $T$, $\mu$ and $m$ axes) \cite{griffiths,rehr}.} Note the appearance of the kinetic term. The sigma field is no longer a constant beyond the mean field approximation. The potential (\ref{true}) will eventually lead to the {\it scaling equation of state} \cite{widom} written in terms of the scaling fields $A_1$ and $A_2$ (the revised scaling \cite{rehr}). Because $T$, $\mu$ and $m$ participate in the magnetic field-like scaling field, we obtain, very schematically, the most singular part \footnote{In calculating $\chi_{ch}$, dominant contribution to $\frac{dA_1}{dm}$ comes from $\sigma_0\frac{d(T-T_c(m))}{dm}$ rather than $\frac{d\sigma_0}{dm}(T-T_c(m))$. The latter term, being proportional to $T-T_c(m)$, behaves as the correction to the scaling. Also, if the derivative acts on $A_2$, we get $\langle\sigma'^2\sigma'^2\rangle\sim |g-g_c|^{-\alpha/\beta \delta}$ which is less singular than the magnetic susceptibility.} 
\begin{eqnarray}
\chi_q &\sim &\frac{\partial^2\Omega}{\partial \mu^2}\sim m^{\frac{2}{5}}\frac{\partial^2 \Omega}{\partial A_1^2}\sim m^{\frac{2}{5}}\langle\sigma'\sigma'\rangle  \nonumber \\
C&\sim & \frac{\partial^2\Omega}{\partial T^2}\sim m^{\frac{2}{5}}\frac{\partial^2 \Omega}{\partial A_1^2}\sim m^{\frac{2}{5}}\langle\sigma'\sigma'\rangle\nonumber \\
\chi_{ch} &\sim &\frac{\partial^2 \Omega}{\partial m^2}\sim m^{-\frac{4}{5}}\frac{\partial^2 \Omega}{\partial A_1^2}\sim
m^{-\frac{4}{5}}\langle\sigma'\sigma'\rangle\nonumber \\
&&\langle\sigma'\sigma'\rangle\sim |g-g_c|^{-\epsilon}
\end{eqnarray}
where $|g-g_c|$ is the distance from the critical end-point in some units. $\epsilon \equiv \gamma/\beta \delta \approx 0.8$ for any direction which make an angle to the $A_1=0$ line at the critical end-point. For the path asymptotically parallel to that line, the exponent is $\gamma\sim 1.2 > \epsilon$. [These values are taken from the 3D Ising model.]

Having discussed the singular behavior of susceptibilities inside the critical region, however, we give a pessimistic result. Since we now have the Landau-Ginzburg potential for CEP, we can say something about the size of the critical region. Recall that, according to the Ginzburg criterion (\ref{tg}), the radius of the critical region is proportional to the square of the coefficient of the quartic term. Other coefficients are quark mass independent in the leading order. Thus we obtain 
\begin{eqnarray}
|t|\sim A_4^2\sim m^{\frac{4}{5}}.\label{dep}
\end{eqnarray}
This gives a bound to the size of the critical region. It shrinks to zero as the quark mass decreases (See, Fig. \ref{phase}).  The physical reason behind this is that the coefficient of the quartic term is zero at the tricritical point and remains small near it.

Generally speaking, the critical point of a strongly interacting system has a large critical region \cite{kogut}. Thus the size of the critical region of CEP is subject to a competition between these opposite effects and the determination of it is a highly nontrivial problem. However, it seems to us that the above bound (\ref{dep}) is a compelling reason to expect that the critical region is `small'.

If the critical region of CEP is small, probably most of the fluctuations associated with CEP come from the mean field region around the critical region.\footnote{It must be cautioned that the mean field region does not always exist. For example, it is known that there is no mean field region for the $\lambda$-transition of liquid helium (the critical region is large, $|t|\sim 0.3$). However, if the critical region is squeezed by an explicit parameter of the theory as in the present situation, it would be meaningful to discuss the mean field region belonging to the critical point. [We thank M. A. Stephanov for a discussion on this point.]}
The central point of this paper is that {\it if we consider the mean field region belonging to CEP, we should also consider the mean field region belonging to TCP.} The tricritical point has, so to speak, a `tricritical region' (see Fig. \ref{phase}) which is a sphere or an ellipsoid in the $(T,\mu,m)$ space centered at $(T_t,\mu_t,0)$.\footnote{Here we use the term `tricritical region' loosely for the region where any mean field-like effects of the tricritical point on susceptibilities exist. This terminology is a bit misleading because there is no critical region for a tricritical point in the usual sense.} Then it is possible that {\it the tricritical region survives in the physical $(T,\mu)$ plane}. The magnitude of the $u$, $d$-quark masses is crucial to this. More interesting possibility is that {\it the critical point is inside the tricritical region and a crossover of different universality classes happens} (not to be confused with the crossover phase transition at lower chemical potentials). Namely, as we approach CEP the critical exponents gradually change from those of the tricritical point to those of the 3D Ising model {\it via} those of CEP in the mean field approximation. [Note that the mean field exponents of a bi-critical point are different from those of a tricritical point.]  Indeed, such kind of crossover was experimentally observed in an antiferromagnet dysprosium aluminium garnet (DAlG) long ago. The critical exponent $\beta$ for the magnetization tends to change from the tricritical value ($\beta=1$) to the Ising model value ($\beta \simeq0.31$) as we go along the wing critical line \cite{giordano}.

Thus, through the consideration of the critical region, we have become aware of a possible interesting role played by the hidden tricritical point. Its critical phenomena are therefore worth studying and will be discussed in the next subsection.

\subsection{The QCD tricritical point}
Motivated by the above arguments, we now turn our attention back to the QCD tricritical point. 
Because the upper critical dimension of models described by (\ref{gl}) is three, the origin of the coupling constant is an attractive IR fixed point. Correspondingly, universal behaviors associated with the tricritical point are well described by the mean field theory up to logarithmic corrections.\footnote{This is why we neglected the pion degrees of freedom in (\ref{gl}). Mean field theory is truly universal in the sense that it does not depend on even the symmetry of the order parameter. However, the multiplicative logarithmic corrections to the scaling do depend on the symmetry of the order parameter.}

Let us see how susceptibilities scale with respect to $|T-T_t|$, $|\mu-\mu_t|$ and $m$ in the mean field approximation. 

At $(T,\mu,m=0)$, straightforward calculations show that
\begin{eqnarray}
\chi_q&\sim &|h-h_t|^{-\gamma_q},\label{gamma}\nonumber \\
\chi_{ch}&\sim &|h-h_t|^{-\gamma_{ch}},
\end{eqnarray}
where $|h-h_t|$ is the distance (in some units) from TCP in the $(T,\mu)$ plane. $\gamma_q=\frac{1}{2}$, $\gamma_{ch}=1$
for paths which are not asymptotically tangential to the first order phase transition line. 

At $(T,\mu,m\neq 0)$,
the expectation value of $\sigma$ is given by the following equation
\begin{eqnarray}
m=a\sigma +b\sigma^3+\sigma^5.
\end{eqnarray}
Near $(T_t, \mu_t, m)$ where $a=b=0$, [Note that this is the `nearest' point to TCP in the phase diagram with a quark mass $m$.] we can expand the solution up to the second order in $a$ and $b$
\begin{eqnarray}
\sigma=m^{\frac{1}{5}}-\frac{a}{5}m^{-\frac{3}{5}}-\frac{b}{5}m^{-\frac{1}{5}}
+O(a^2m^{-\frac{7}{5}},b^2m^{-\frac{3}{5}},abm^{-1}).\label{kai}
\end{eqnarray}
Inserting (\ref{kai}) into (\ref{gl}) and differentiating with respect to $\mu$ twice, we get $\chi_q$. Because of (\ref{ab}) the differentiation with respect to $\mu$ is replaced by the differentiation with respect to $a$ and $b$. Extracting the most singular contribution, we obtain
\begin{eqnarray}
\chi_q\sim \frac{\partial^2 \Omega[\sigma]}{\partial \mu^2}|_{a=b=0} \sim m^{-\frac{2}{5}}.\label{h}
\end{eqnarray}
Analogously, 
\begin{eqnarray}
\chi_{ch}\sim m^{-\frac{4}{5}}.
\end{eqnarray}
The divergence of $\chi_q$ is rather moderate in the mass direction, from which we expect that the quark number susceptibility may still be large even with non-zero quark masses. Indeed, from (\ref{cc}) and (\ref{dep}) we can derive the $m$-dependence of $\chi_q$ at the edge of the critical region
\begin{eqnarray}
\chi_q\sim m^{\frac{2}{15}}|t|^{-\frac{2}{3}}\sim m^{\frac{2}{15}}(m^{\frac{4}{5}})^{-\frac{2}{3}} \sim m^{-\frac{2}{5}}.
\end{eqnarray}
Comparing with (\ref{h}), we see that the $m$-dependence is exactly the same. There may or may not be a reason for this coincidence. In any case, this does show that TCP is as important as CEP at least in the small quark mass limit. \\

Starting from the simple Landau-Ginzburg potential, we have extracted a lot of physics near CEP/TCP. These analyses show the power of universality as well as its limitations. For example, the universality argument does not tell us whether the effect of TCP survives in the ($T, \mu$) plane with the quark mass of, say, 5 MeV.
In order to quantify the ideas given in this section, we must resort to a specific model. This will be the subject of the next section.

\section{Numerical results}

 In this section, we numerically calculate the quark number susceptibility in the ($T, \mu$) plane by using a model. As expected, the susceptibility diverges both at the critical and tricritical points. We also calculate the corresponding critical exponent. The results clearly demonstrate that the hidden tricritical point can affect the phase diagram with nonzero quark masses.

\subsection{CJT effective potential and the chiral phase transition}

As a model, we employ the Cornwall-Jackiw-Tomboulis (CJT) effective potential
\cite{CJT}
for 2-flavor QCD in the improved ladder approximation 
\cite{Kiriyama:2001ah}. A brief description of the model is given in Appendix A. For more details, see \cite{Kiriyama:2001ah}.

At zero temperature and chemical potential, the effective potential $V$ is given by
\be
V[\Sigma] = - 2 \int \frac{d^4p}{(2\pi)^4} \ln
              \frac{\Sigma^2(p) + p^2}{p^2} 
            - \frac{2}{3 C_F} \int dp^2 \frac{1}
                 {\frac{d}{dp^2} \frac{\bar{g}^2(p^2)}{p^2}   }
                \left( \frac{d \Sigma(p)}{dp^2} \right) \, ,
\label{eq:T0CJT}
\ee
where the gauge coupling constant $\bar{g}^2(p^2)$ and 
the dynamical quark mass function $\Sigma(p)$ are
\beq
\bar{g}^2(p^2) &=& \frac{2\pi^2 a}{\ln[(p^2+p^2_c)/\La^2]}
         \, , 
\label{eq:T0coupling}\\
\Sigma(p) &=& m_q \left( \ln[(p^2+p^2_c)/\La^2] \right)^{-a/2} +
            \frac{\sigma}{p^2+p_c^2} \, 
            \left( \ln[(p^2+p^2_c)/\La^2] \right)^{a/2-1} \, .
\label{eq:T0mass}
\eeq
$p_c$ is a momentum scale which separates the
infrared (nonperturbative) region from the ultraviolet (perturbative) region. 
$C_F = (N_c^2-1)/2N_c$ is the quadratic 
Casimir operator for the fundamental representation of the color SU($N_c$) group and $a \equiv 24/(11N_c-2N_f)$ ($N_f$ is the number of active flavors.\footnote{Although the potential is evaluated with $N_f=2$, we take $N_f=3$ in the gauge coupling (\ref{eq:T0coupling}). In this way we include the effect of $s$-quark only through the running of the coupling constant.} 
$\sigma$ is proportional to the renormalization group invariant
chiral condensate $\CC$ as $\sigma = 2\pi^2a\CC/3$ and
$m_q$ is the  renormalization group invariant current quark mass.
They are related to the scale dependent mass 
$m_q^{\Lambda}$ and the scale dependent chiral condensate $\CC^{\Lambda}$
by the perturbative renormalization group equation
\beq
          \CC &=& \frac{ \CC_{\Lambda}}
                      {\left[ \ln(\Lambda^2/\La^2) \right]^{a/2}} 
          \, ,\label{eq:scale_CC}\\
          m_q &=& m_q^{\Lambda} 
          \left[ \ln(\Lambda^2/\La^2) \right]^{a/2} \, .
          \label{eq:scale_m}
\eeq

An overall factor ($N_f=2$ times $N_c=3$)
is omitted in Eq. (\ref{eq:T0CJT}). The chiral condensate $\CC$ and $f_{\pi}$ are known to be insensitive to 
the infrared regularization parameter $p_c$ 
\cite{Kugo:1992zg}.
Therefore we take $p_c^2/\La^2=e^{0.1}$ and determine $\La$ to 
reproduce the pion decay constant $f_{\pi}=93$ MeV in the Pagels-Stokar formula
\cite{PS} in the chiral limit. We obtain 
$\La = 738$ MeV for $N_f = 2$
\cite{Kiriyama:2001ah}.
In the following calculations, we take $\Lambda=1$ GeV in (\ref{eq:scale_m}) and change the value of  $m_q^{\Lambda=1 \mbox{\scriptsize{GeV}}}$. For simplicity, we abbreviate $m_q^{\Lambda=1 \mbox{\scriptsize{GeV}}}$ to $m_q$ below.

At finite temperature and chemical potential,
we use the imaginary time formalism 
\cite{ITF},
and make the replacement 
\be
\int \frac{d^4p}{(2\pi)^4} f(\p,p_4) \rightarrow
T \sum_{n=-\infty}^{\infty} \int \frac{d^3\p}{(2\pi)^3}
f(\p,\wn + i \mu) \, ,
\label{eq:imaginary}
\ee 
where $\wn = (2n+1)\pi T \, (n \in Z)$ is the Matsubara-frequency for the quark. \footnote{However, we replace $p_4$ with $\omega_n$, not $\omega_n+i\mu$ in the gauge coupling (\ref{eq:T0coupling}) to avoid an absurd situation.} 

As a normalization, we define $\tilde{V}$ by subtracting the $\sigma$-independent part from $V$ such that $\tilde{V}$ reduces to the value of the free quark gas when $\sigma=0$. See Appendix A.


We can study the chiral phase transition and the phase diagram
by calculating $\tilde{V}[\sigma,m_q]$ at given $T$ and $\mu$ and searching the value of the chiral condensate $\sigma_0$ which minimizes the potential.
The location of the first order phase transition line is determined by finding a gap in $\sigma_0$.  In the chiral limit, $\sigma_0$ goes to zero smoothly as the second order phase transition line is approached from below.
With finite quark masses, there is no distinct border
between the symmetric and broken phases, and $\sigma_0$ remains finite at all temperatures and chemical potentials.

\begin{figure}[htbp]
    \centerline{
      \epsfxsize=0.49\textwidth
      \epsfbox{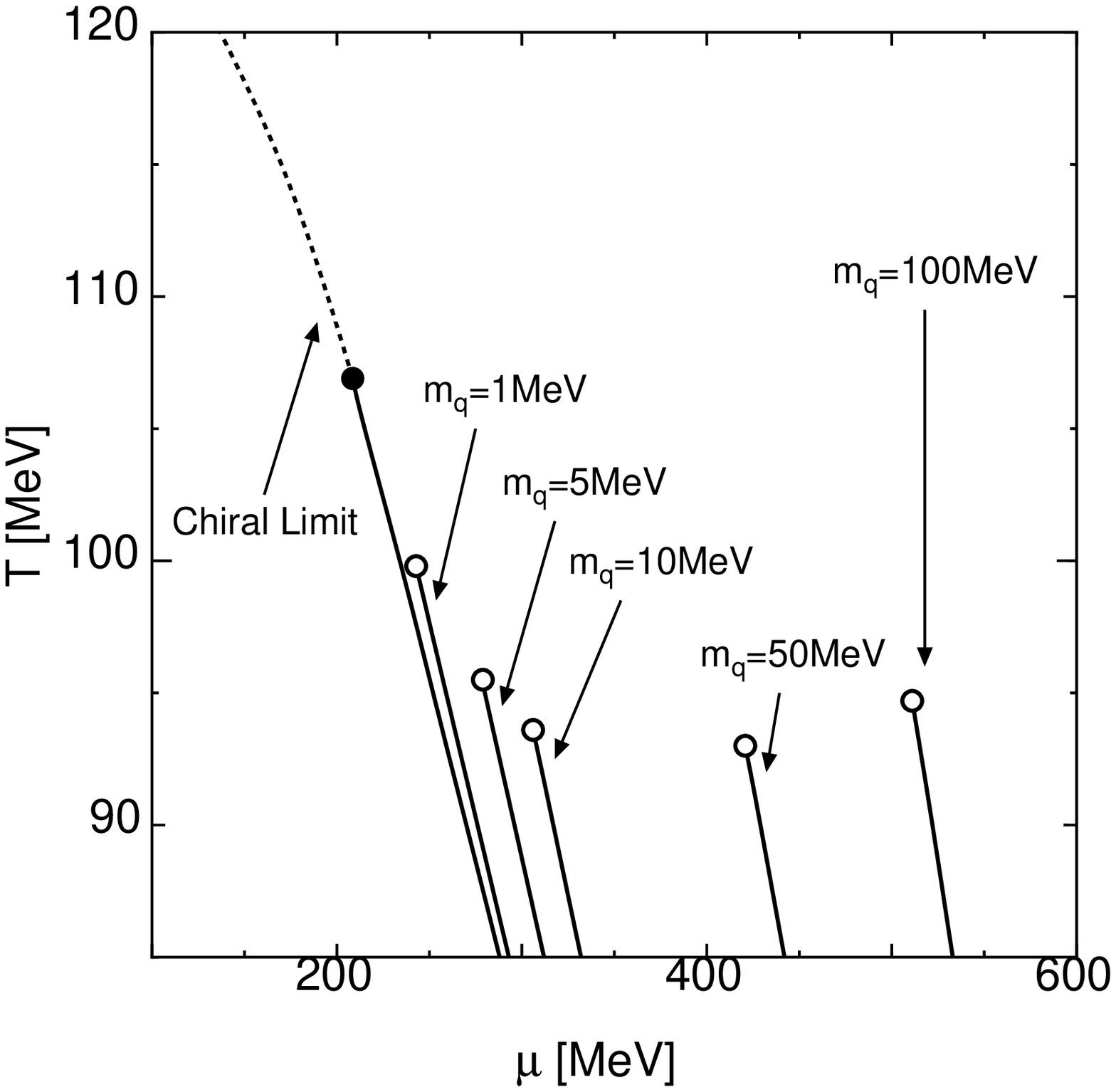}
    }
  \caption{The phase diagram with several quark masses. The quark
           masses are evaluated at the momentum scale 1 GeV.
           The solid
           and dotted lines represent the first order and the second
           order phase transition, respectively. The filled circle is the tricritical point and open circles are the critical end points 
           for different quark masses.}
  \label{fig:chiral_transition}
\end{figure}

\begin{figure}[htbp]
    \centerline{
      \epsfxsize=0.49\textwidth
      \epsfbox{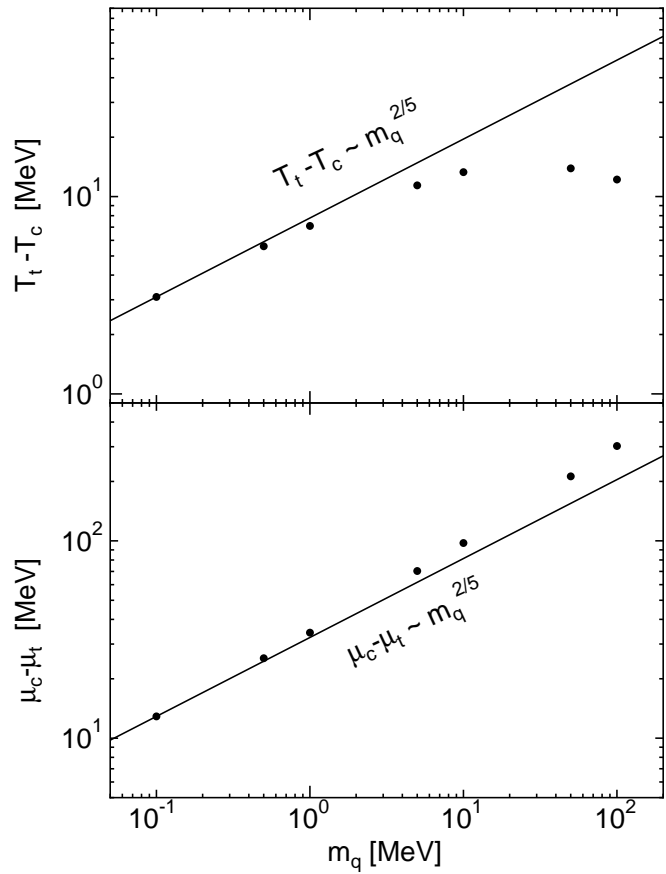}
    }
  \caption{The quark mass dependence of the critical temperature (upper figure)
           and the critical chemical potential (lower figure). The slope of the solid line is $\frac{2}{5}$.}
  \label{fig:CEP_m_dep}
\end{figure}

The phase diagram with several quark masses in the ($T,\mu$) plane 
is shown in Fig. \ref{fig:chiral_transition}.
The location of the tricritical point in the chiral limit
is $T_t = 107$ MeV and 
$\mu_t= 209$ MeV.
The open circles
in Fig. \ref{fig:chiral_transition} represent the critical end points
for different quark masses. 
As shown in Fig. \ref{fig:CEP_m_dep}, the distance between TCP and CEP approximately scales as $m_q^{2/5}$ up to $m_q\sim O(1)$ MeV, in agreement with (\ref{mm}). 
For larger masses, $m_q >$ 10 MeV, $T_c(m_q)$ does not change much
while $\mu_c(m_q)$ keeps on 
increasing.

\subsection{The quark number susceptibility around CEP/TCP}

\begin{figure}[htbp]
    \centerline{
      \epsfxsize=0.49\textwidth
      \epsfbox{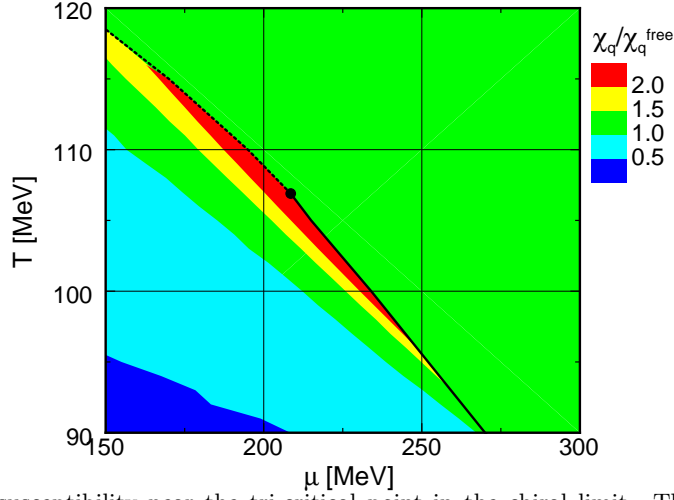}
    }
  \caption{The quark number susceptibility near the tri-critical point
           in the chiral limit. The value of the 
           susceptibility is divided by
           that of the massless free quark gas. The solid and dotted
           line represent the first and the second order phase transition, respectively and the filled circle is the tricritical point.}
  \label{fig:chiral}
\end{figure}

\begin{figure}[htbp]
    \centerline{
      \epsfxsize=0.49\textwidth
      \epsfbox{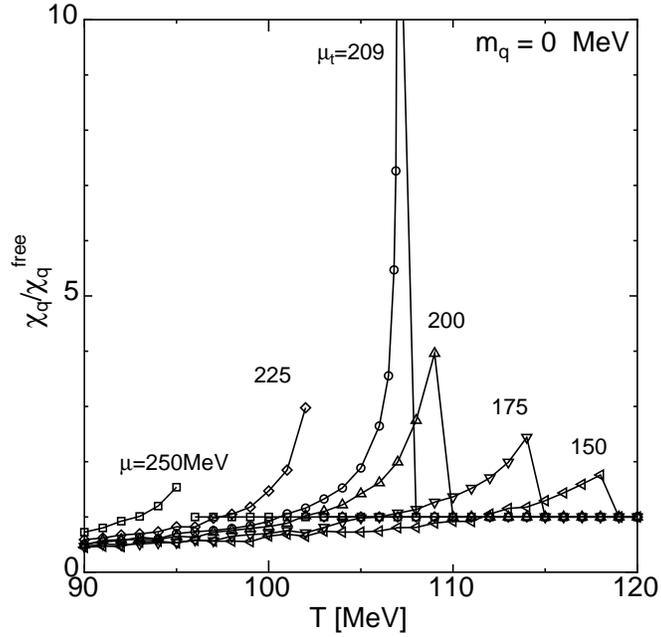}
    }
  \caption{The temperature dependence of $\chi_q$ at fixed $\mu$'s. For $\mu<209$ MeV, $\chi_q$ has a jump across the second order phase transition line (O(4) line), which is consistent with the mean field theory. See Appendix B.
}
  \label{mono}
\end{figure}

The quark number susceptibility $\chi_q$ is calculated
from the normalized effective potential $\tilde{V}$ as
\be
\chi_q = - \frac{\partial^2 \tilde{V}[\sigma_0]}{\partial \mu^2} \, .
\label{eq:chi_q}
\ee
Fig. \ref{fig:chiral} and Fig. \ref{mono} are the results in the chiral limit. 
As we can see, $\chi_q$ is suppressed far below the chiral phase 
transition line and 
enhanced near TCP.
In the chirally symmetric phase, $\chi_q $ is equal to the value of the massless free quark gas $\chi_q^{\mbox{\scriptsize free}}$
in this model. The region where $\chi_q$ is enhanced is elongated in the direction parallel to the first order phase transition line. This is because the critical exponent for this direction ($\gamma_q=1$) is larger than for other directions ($\gamma_q=\frac{1}{2}$).
We also found a jump in $\chi_q$ 
along the second order phase transition line.
Inside the critical region, however, the jump must be replaced by a {\it cusp} with certain critical exponents. See Appendix B. Our model can reproduce only the mean field behaviors.

\vspace{1cm}

\begin{figure}[htbp]
    \centerline{
      \epsfxsize=0.49\textwidth
      \epsfbox{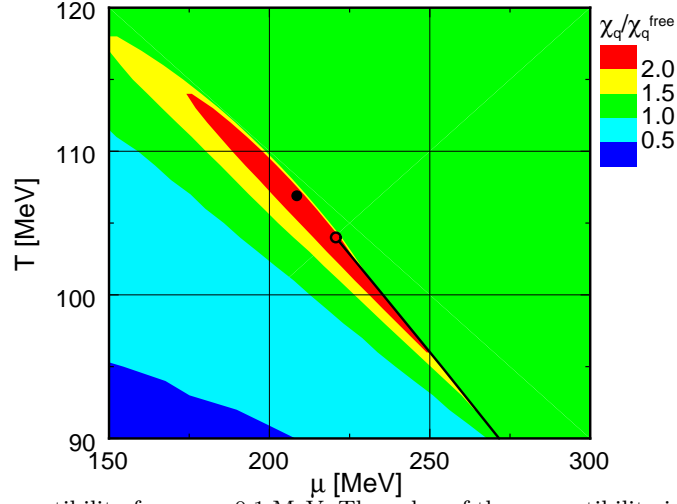}
    }
  \caption{The quark number susceptibility for 
           $m_q = 0.1$ MeV.
           The value of the 
           susceptibility is divided by
           that of the massless free theory. 
           The solid line is the first order transition line.
           The open circle represent the critical end point
           for $m_q = 0.1$ MeV.
           The filled circle is at ($T_t, \mu_t$).
           }
  \label{fig:chi_mq01}
\end{figure}

\begin{figure}[htbp]
    \centerline{
      \epsfxsize=0.49\textwidth
      \epsfbox{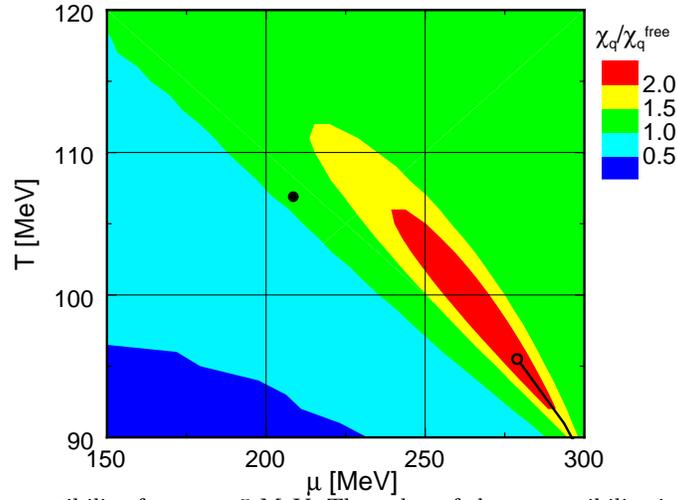}
    }
  \caption{The quark number susceptibility for 
           $m_q = 5$ MeV. The value of the 
           susceptibility is divided by
           that of the massless free theory. 
           The open circle is the critical end point
           for $m_q = 5$ MeV and the filled circle is at ($T_t, \mu_t$).
           }
  \label{fig:chi_mq5}
\end{figure}

Next we examine $\chi_q$ for finite quark masses. 
Fig.\ref{fig:chi_mq01} and Fig. \ref{fig:chi_mq5}. are the results for 
$m_q = 0.1$ MeV and $m_q=5.0$ MeV, respectively.
The location of CEP is ($T_c, \mu_c$)=(104 MeV, 221 MeV) for $m_q=0.1$ MeV 
and (95 MeV, 279 MeV) for $m_q=5.0$ MeV. $\chi_q$ diverges at CEP and 
is enhanced in the elongated region parallel to the first order phase 
transition line because the critical exponent is the largest 
for this direction as in the massless case. 
For $m_q=0.1$ MeV, TCP is still close to CEP and the elongated region includes the point ($T_t, \mu_t$) while for  $m_q=5.0$ MeV, the region deviates from it.

At first sight, one might think that the analysis made in the previous section ceases to be valid at somewhere between $m_q=0.1$ MeV and $m_q=5$ MeV and the effect of TCP no longer survives for $m_q=5$ MeV which might be considered as the `realistic' quark mass in this model. \footnote{In this model
          $\CC_{\Lambda=\mbox{\scriptsize 1GeV}}=(-276\mbox{MeV})^3$ at $T=\mu=0$. By using Gell-Mann-Oakes-Renner relation with $m_{\pi}=140$ MeV,
          $m_q^{\Lambda=\mbox{\scriptsize 1GeV}} \sim 4$MeV.} 
However, this conclusion is too hasty. We will see in the next subsection that the hidden tricritical point still affects the physics near CEP even for $m_q=5$ MeV.

\vspace{1.5cm}

\vspace{1.5cm}

\subsection{The critical exponent for $\chi_q$}

Now let us examine the critical exponent for $\chi_q$ 
at CEP and TCP. We calculate it along the path parallel
to the $\mu$ axis in the $T$-$\mu$ plane from lower $\mu$ towards CEP/TCP
at fixed $T_c$ or $T_t$.

\begin{figure}[htbp]
    \centerline{
      \epsfxsize=0.49\textwidth
      \epsfbox{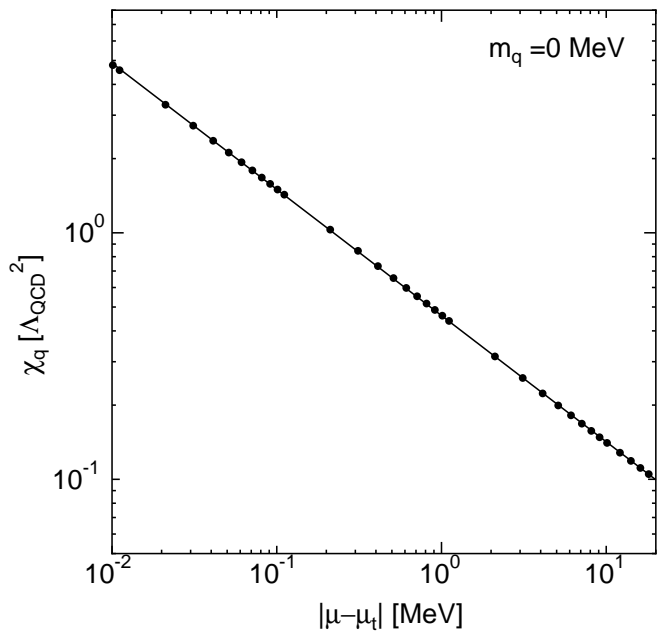}
    }
  \caption{The quark number susceptibility in the chiral limit
           as a function of $|\mu - \mu_t|$ at fixed temperature $T_t$.
          } 
  \label{fig:chiral_expo}
\end{figure}

First we consider the chiral limit. 
We expand $\tilde{V}$ in the vicinity of TCP.\footnote{The reason for this expansion is twofold. First, in order to keep in line with the argument given in Section II. Second, technically we can approach TCP much closer to determine the exponent than directly reading it from Fig. \ref{fig:chiral}.}
\be
\tilde{V}[\sigma,m_q=0] = V_{\mbox{\scriptsize free}} + a_2(T,\mu)\sigma^2
 + a_4(T,\mu)\sigma^4 + a_6(T,\mu)\sigma^6 \, .
\label{eq:expansion}
\ee
The coefficients $a_2,a_4,a_6$ and 
$V_{\mbox{\scriptsize free}}$ are summarized in Appendix A.
$\sigma_0$ is determined by the equation
$\left. \partial \tilde{V}/\partial \sigma \right|_{\sigma = \sigma_0} = 0$,
We obtain
\be
\tilde{V}[\sigma_0 = 0,m_q=0] = V_{\mbox{\scriptsize free}} 
\, ,
\label{eq:V_W} \\
\ee
above the chiral transition line, and
\be
\tilde{V}[\sigma_0,m_q=0] = V_{\mbox{\scriptsize free}}
+\frac{a_4}{27a_6^2}(2a_4^2-9a_2a_6)-\frac{2}{27a_6^2}(a_4^2-3a_2a_6)^{3/2}
\, ,
\label{eq:V_NG}
\ee
below that line. $\chi_q$ is obtained by taking the second derivative of
(\ref{eq:V_W}) and (\ref{eq:V_NG}) with respect to $\mu$.
Fig. \ref{fig:chiral_expo} shows 
$\chi_q$ for numbers of $|\mu-\mu_t|$'s. We determine the critical 
exponent $\gamma_q$ defined in (\ref{gamma})
numerically by using a linear logarithmic fitting
\be
\ln \chi_q = -\gamma_q \ln|\mu-\mu_t| 
+ \mbox{const}. \, ,
\ee
where const. is independent of $\mu$.
We obtained $\gamma_q = 0.51 \pm 0.01$ which is consistent with the mean field theory.

\begin{figure}[htbp]
    \centerline{
      \epsfxsize=0.49\textwidth
      \epsfbox{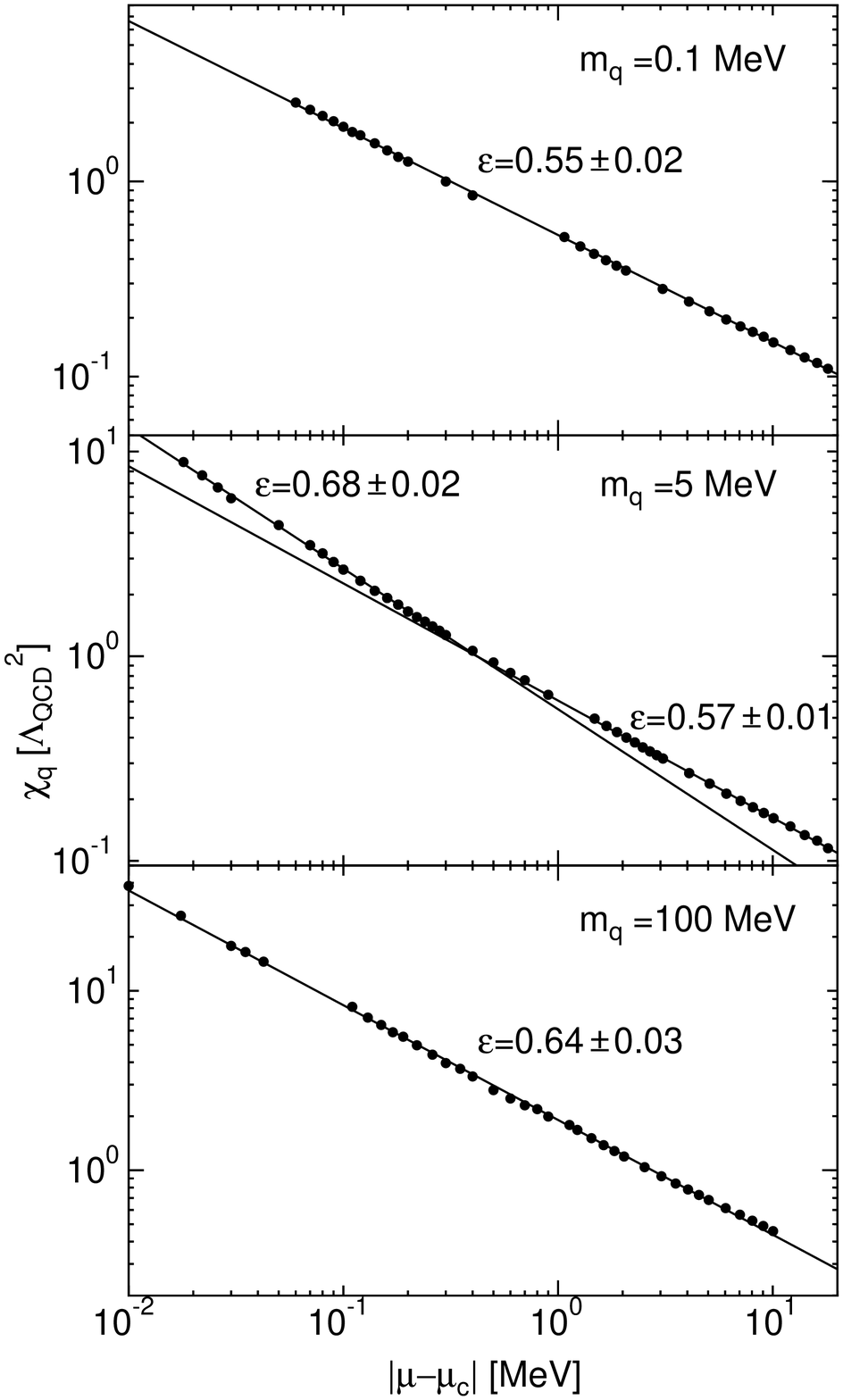}
    }
  \caption{The quark number susceptibility for $m_q$ = 0.1, 5 and 100 MeV as a function of $|\mu-\mu_c|$ at fixed temperature $T_c(m_q)$.
          } 
  \label{fig:finite_expo}
\end{figure}

With finite quark masses, the expectation value $\sigma_0$ is determined only numerically. This time we do not expand the potential around $\sigma_0$ and directly read the exponent from Fig. \ref{fig:chi_mq01} and Fig. \ref{fig:chi_mq5}. In Fig. \ref{fig:finite_expo}, 
$\chi_q$ is plotted for numbers of $|\mu-\mu_c|$'s
for $m_q$ = 0.1, 5, and 100 MeV together with the calculated values of the critical exponent $\epsilon$ defined in (\ref{cc}).
\be
\ln \chi_q = -\epsilon \ln|\mu-\mu_c| 
+ \mbox{const}. \, .
\ee
For $m_q$ = 0.1 MeV we obtained $\epsilon =0.55 \pm 0.02$.
This is significantly different from the prediction of the mean field theory $\epsilon = \frac{2}{3}$, which is a clear evidence of the effect of the tricritical point. We expect that the exponent changes towards $\frac{2}{3}$ if we approach CEP much closer.

For $m_q$= 5 MeV, the slope of the data points 
changes at around $|\mu-\mu_c| \sim 0.5$ MeV.
Therefore we fitted the data for $|\mu-\mu_c| < 0.3$ MeV 
and $> 1$ MeV separately and obtained the critical exponent
$ 0.68 \pm 0.02 $ for $|\mu-\mu_c| < 0.3$ MeV and $0.57 \pm 0.01$
for $|\mu-\mu_c| > 1$ MeV. We interpret this change of the exponent as the {\it crossover} of different universality classes discussed in the previous section. Note that the purely mean field-like exponent is seen in a very small region $|\mu-\mu_c| < 1$ MeV from CEP. This result is somewhat surprising to the present authors because, as seen in Fig. \ref{fig:chi_mq5}, TCP is far away from CEP already for  $m_q$= 5 MeV and the value of $\chi_q$ itself is unremarkable at ($T_t, \mu_t$). 
It seems that, although the analysis in the previous section was made in the small quark mass limit, the effect of TCP is unexpectedly robust against the increase of the quark mass.   

As a check, we also calculated the exponent for $m_q=$ 100 MeV and obtained $\epsilon =0.64 \pm 0.03$ which is consistent with the mean field value $\frac{2}{3}$.
For such a large quark mass, we see no indication of a change in the slope. The effect of TCP has completely disappeared.

\section{conclusions}
Based on the universality argument and numerical model calculations, we studied the singular behavior of susceptibilities near the critical/tricritical points. These two approaches are complementary, and we observed that the model calculation faithfully quantified the qualitative predictions obtained by the universality argument as long as the mean field behaviors are concerned. The important point is that, although we adopted a specific model, the qualitative behavior of $\chi_q$ is probably model independent.
In particular, our results strongly suggest a possibility that the tricritical point affects the physics near the critical end-point. In other words, there are {\it traces} of the hidden tricritical point on the QCD phase diagram.
Practically, the traces will be seen as the gradual change of the critical exponents since, after all, universality classes are characterized only by their critical exponents. It is expected that the exponents change from those of TCP to those of the Ising model via those of CEP in the mean field approximation. In order to really confirm this fascinating possibility, lattice simulations at finite chemical potentials \cite{fodor} are necessary.

Finally, we briefly comment on the implication of our results to heavy-ion experiments. The divergence of $\chi_q$ is directly related to an anomaly in the event-by-event fluctuation of baryon number $B$ (divided by the entropy $S$)
\begin{eqnarray}
\frac{\langle (\Delta B)^2\rangle}{S},
\end{eqnarray}
which was originally introduced in \cite{asakawa} to probe the deconfined phase. Although neutrons are not observed, we expect that the event-by-event fluctuation of the {\it proton number} is relatively enhanced for collisions which have passed `near' CEP/TCP. Pion and diphoton observables are discussed in \cite{Stephanov:1998dy,Berdnikov:1999ph,fukushima}. As we remarked before, the critical exponents of the Ising model and the mean field theory are not so different numerically. Thus, the smallness of the critical region itself may not be an obstacle to the observability of critical phenomena in experiments. However, if we take the effect of TCP seriously either by assumption or stimulated by future lattice results, we must take into account the long-wavelength fluctuations of the {\it pions} as well as the sigma meson because the pions are no longer the `environment' but participate in the critical fluctuations around the trace of TCP.

\section*{ACKNOWLEDGMENTS}

We greatly thank T. Hatsuda and T. Kunihiro for their continuous encouragement 
and numerous valuable discussions. We also thank M. Asakawa, K. Itakura, L. McLerran, R. D. Pisarski and M. A. Stephanov for discussions and comments. 
T. I. is supported by Special Postdoctral Researchers Program 
of RIKEN. Y. H. acknowledges RIKEN BNL Research Center where this work was completed.


\appendix

\section{Description of the model}

\subsection{The normalized CJT effective potential}

We begin with the Cornwall-Jackiw-Tomboulis(CJT) effective potential
\cite{CJT}
for QCD in the improved ladder approximation
\cite{Kiriyama:2001ah}
as a functional of the quark propagator $S(p)$
at zero temperature and quark chemical potential
after the Wick rotation,
\beq
V[S] &=& V_1[S] + V_2[S] ,\\
V_1[S] &=& \int \frac{d^4p}{(2\pi)^4} \mbox{tr} \left\{ \ln \left[
       S_0^{-1}(p)S(p) \right] - S_0^{-1}(p)S(p) + 1 \right\} ,\\
V_2[S] &=& - \frac{1}{2} \int \int
             \frac{d^4p}{(2\pi)^4} \frac{d^4k}{(2\pi)^4} g^2(p-k)
       \left\{ \mbox{tr}
       \left[ \frac{\lambda^a}{2} \gmu S(k)\frac{\lambda^a}{2} \gnu S(p) 
       \right]  D_{\mu\nu}(p-k) \right\} .
\eeq
Here ``tr'' is taken over the Dirac, flavor and color matrices
(Gell-Mann matrices $\lambda^a$), and
$S_0(p)$ and $D_{\mu\nu}(p-k)$ are the free quark propagator
and the gluon propagator in the Landau gauge ($D_{\mu\nu}(p-k) = (\delta_{\mu\nu} - p_{\mu}p_{\nu}/p^2)/p^2$), respectively.
$V_1[S]$ corresponds to the 1-loop potential with 
the quark 1-loop diagram and
$V_2[S]$ is the 2-loop potential with the one gluon exchange.

We adopt the following approximation, the so-called Higashijima-Miransky approximation 
\cite{Higashijima,Miransky}
for the QCD running coupling constant
\be
g^2\left( (p-k)^2 \right) 
\rightarrow \theta(p^2-k^2)\bar{g}^2(p^2) + \theta(k^2-p^2)\bar{g}^2(k^2) 
         \, ,
\ee
where $\bar{g}$ is defined in (\ref{eq:T0coupling}).
In this approximation with the Landau gauge, the renormalization
of the quark wave function may be neglected at zero temperature and chemical potential. At finite temperature and chemical potential we need to take the wave function renormalization into account even in the Landau gauge \cite{Ikeda:2001vc}. However, we ignore this problem for the present purpose.
Then the CJT effective potential can be rewritten as (\ref{eq:T0CJT}) in terms of
the dynamical quark mass function $\Sigma(p)$ using the 
corresponding Schwinger-Dyson equation for $\Sigma(p)$.

As a normalization, we define $\tilde{V}$ by subtracting the $\sigma$-independent part from $V$ such that $\tilde{V}$ reduces to the value of the free quark gas when $\sigma=0$. We obtain
\beq
\tilde{V}[\sigma,m_q] =V_{\mbox{\scriptsize free}} &-& \frac{T}{\pi^2} \sum_{n=0}^{\infty}
                  \int_0^{\infty} d|\p| \p^2 
                  \ln \frac{(\Sigma^2(\p^2,\wn^2;\sigma,m_q) 
                             +\p^2+\wn^2-\mu^2)^2
                            + 4\mu^2\wn^2}
                           {(\Sigma^2(\p^2,\wn^2;0,m_q) +
                            \p^2+\wn^2-\mu^2)^2+ 4\mu^2\wn^2} \nonumber \\
                &+& \frac{16T}{3 C_F a \pi^2} \sum_{n=0}^{\infty}
                  \int_0^{\infty} d|\p| \p^2
                  \frac{(\p^2+\wn^2)(\p^2+\wn^2+p^2_c)\left[\ln
                        (\p^2+\wn^2+p^2_c) \right]^2}
                       {(\p^2+\wn^2+p^2_c) \ln(\p^2+\wn^2+p^2_c)
                        + \p^2 +\wn^2}  \nonumber \\
                & &\times
                        \left( m_q a \sigma \frac{\left[ \ln
                        (\p^2+\wn^2+p^2_c) \right]^{-2}}
                        {(\p^2+\wn^2+p^2_c)^3} 
                        \left[ \ln(\p^2+\wn^2+p^2_c)+1-\frac{a}{2} \right]
                        \right.
                  \nonumber \\
                & & \hspace{20pt} \left. - \frac{\sigma^2}
                        {(\p^2+\wn^2+p^2_c)^4} \left[ \ln
                        (\p^2+\wn^2+p^2_c) \right]^{a-4} \,
                        \left[ \ln(\p^2+\wn^2+p^2_c)+1-\frac{a}{2} \right]^2
                         \right),
\label{eq:renorm_CJT}
\eeq
where $a=24/(11N_c-2N_f)$ and the effective potential for the free quark 
$V_{\mbox{\scriptsize free}}$ is given by
\be
V_{\mbox{\scriptsize free}} = - 2 T \int \frac{d^3\p}{(2\pi)^3} \left[
           \ln \left(1+ e^{-(\omega-\mu)/T} \right) +
           \ln \left(1+ e^{-(\omega+\mu)/T} \right) \right]
\ee
with $\omega = \sqrt{\p^2 + m_q^2}$.
In the chiral limit ($m_q = 0$),the momentum integral can be easily 
performed and $V_{\mbox{\scriptsize free}}$ 
becomes
\be
V_{\mbox{\scriptsize free}}(m_q=0) = -\left( \frac{\mu^4}{12\pi^2} + 
                        \frac{\mu^2T^2}{6} + \frac{7\pi^2T^4}{180} \right).
\label{eq:chiral_p}
\ee
The quark number susceptibility of the massless free quark gas is given by
(omitting the overall factor $N_f N_c$) 
\be
\chi_q^{\mbox{\scriptsize free}}(m_q=0) = \frac{T^2}{3} + \frac{\mu^2}{\pi^2}
\, .
\ee

\subsection{The pion decay constant in the Pagels-Stokar formula}

The parameters $p_c$ and $\La$ are determined such that they reproduce
the pion decay constant $f_{\pi}=93$ MeV in the chiral limit.
We calculate $f_{\pi}$ by the Pagels-Stokar formula
\cite{PS}
\be
f_{\pi}^2 = 4 N_c 
            \int \frac{d^4p}{(2\pi)^4} 
            \frac{\Sigma(p;\sigma_0,m_q=0)}{\left(
            \Sigma^2(p;\sigma_0,m_q=0) + p^2 \right)^2}
            \left[\Sigma(p;\sigma_0,m_q=0) - \frac{p^2}{2}
            \frac{d\Sigma(p;\sigma_0,m_q=0)}{dp^2} \right] \, .
\label{eq:f_pi}
\ee
In above equation, we set $N_f = 2$ because the pion
consists of $u$ and $d$ quarks.

\subsection{Coefficients in the mean field expansion}

The explicit expressions of the coefficients 
$a_2,a_4$ and $a_6$ in Eq. (\ref{eq:expansion}) are
\beq
a_2(T,\mu) &=& \frac{1}{\pi^2} T \sum_{n=0}^{\infty}
               \int d|\p| \p^2 \left\{ - \frac{2 \left[\ln(\p^2+\wn^2+p^2_c)
               \right]^{a-2} (\p^2+\wn^2-\mu^2)}{(\p^2+\wn^2+p^2_c)^2
               \left[(\p^2+\wn^2-\mu^2)^2+4\mu^2\wn^2\right]} \right.
               \nonumber \\
           && +\frac{9}{2}\frac{\left[\ln(\p^2+\wn^2+p^2_c)+1-\frac{a}{2}
              \right]^2 (\p^2+\wn^2) \left[\ln(\p^2+\wn^2+p^2_c)
               \right]^{a-2}}{(\p^2+\wn^2+p^2_c)\left[(\p^2+\wn^2+p^2_c)
              \ln(\p^2+\wn^2+p^2_c) + \p^2+\wn^2\right]} \, , \\
a_4(T,\mu) &=& \frac{1}{\pi^2} T \sum_{n=0}^{\infty}
               \int d|\p| \p^2 \frac{\left[\ln(\p^2+\wn^2+p^2_c)
               \right]^{2a-4} \left[ (\p^2+\wn^2-\mu^2)^2 -4\mu^2\wn^2
               \right]}{(\p^2+\wn^2+p^2_c)^4
               \left[(\p^2+\wn^2-\mu^2)^2+4\mu^2\wn^2\right]^2} \, , \\
a_6(T,\mu) &=& -\frac{2}{3\pi^2} T \sum_{n=0}^{\infty}
               \int d|\p| \p^2 \frac{\left[\ln(\p^2+\wn^2+p^2_c)
               \right]^{3a-6} (\p^2+\wn^2-\mu^2)
               \left[ (\p^2+\wn^2-\mu^2)^2 -12\mu^2\wn^2
               \right]}{(\p^2+\wn^2+p^2_c)^6
               \left[(\p^2+\wn^2-\mu^2)^2+4\mu^2\wn^2\right]^3} \, .
\eeq

\section{The O(4) critical line}

In this appendix, for completeness, we examine the singular behavior of $\chi_q$ along the O(4) line emerging from TCP toward the temperature axis in the $m=0$ plane (See, Fig. \ref{phase}). We call this line the O(4) line because it consists of a sequence of critical points whose universality class is the same as that of the O(4) spin model \cite{rob}. We again start with (\ref{gl}) with $m=0$ and the replacement $\sigma^2 \to \phi^2\equiv \sigma^2+(\pi^1)^2+(\pi^2)^2+(\pi^3)^2 $. The O(4) line in the $(T, \mu)$ plane is determined by the following equation
\begin{eqnarray}
a(T,\mu)=0.
\end{eqnarray}
 Since $b>0$ does not vanish and smoothly varies along this line, we can drop the $\phi^6$ term. If we consider the mean field behavior, we can expand $a$ around an arbitrary point $(T_c,\mu_c)$ on the line \cite{valid}
\begin{eqnarray}
a(T,\mu)=C'(T-T_c)+D'(\mu-\mu_c).\label{1}
\end{eqnarray}
In the mean field approximation, the (singular part of) thermodynamic potential becomes 
\begin{eqnarray}
\Omega_{MF}=0,
\end{eqnarray}
above the O(4) line, and 
\begin{eqnarray}
\Omega_{MF}=-\frac{a^2}{4b},
\end{eqnarray}
below the O(4) line.
Taking the second derivative in $\mu$, we see that the quark number susceptibility has a discontinuous jump across $(T_c,\mu_c)$ and that it is larger in the low temperature phase (below the O(4) line) than in the high temperature phase (above the O(4) line) except for points where $D'=0$.
Beyond the mean field approximation, we use the current theoretical estimate of the specific heat exponent of the O(4) spin model \cite{zinn}
\begin{eqnarray}
\alpha \sim -0.2.
\end{eqnarray}
The minus sign means that the quark number susceptibility shows a cusp at $T_c$ as in the case of the $\lambda$-point of liquid helium. [$\alpha$ is also negative for the O(2) model.] Note that $(T_c,\mu=0)$ is {\it the} point where $D'=0$. It was shown in \cite{allton} that the O(4) line is perpendicular to the temperature axis. Thus the quark number susceptibility has no singularity at ($T_c, \mu=0$) even in the chiral limit and increases monotonously as a function of the temperature, consistent with the results of lattice simulations. However, this smooth behavior is an exception only at $\mu=0$. At any nonzero $\mu$, $\chi_q$ has a cusp precisely at the critical temperature $T_c(\mu)$. The cusp becomes higher and higher as we increase $\mu$ and finally diverges at the tricritical point.


%
 
%

%
%


\begin{references}
\bibitem{valid} This is valid only in the framework of the mean field theory.
\end{references}
\end{document}